# THE INVARIANT MANIFOLD APPROACH APPLIED TO NONLINEAR DYNAMICS OF A ROTOR-BEARING SYSTEM.

Cristiano VIANA SERRA VILLA, Jean-Jacques SINOU* and Fabrice THOUVEREZ

Laboratoire de Tribologie et Dynamique des Systèmes UMR CNRS 5513
Ecole Centrale de Lyon, 3- avenue Guy de Collongues, 69134 Ecully, France.

ABSTRACT

The invariant manifold approach is used to explore the dynamics of a nonlinear rotor, by determining the nonlinear normal modes, constructing a reduced order model and evaluating its performance in the case of response to an initial condition. The procedure to determine the approximation of the invariant manifolds is discussed and a strategy to retain the speed dependent effects on the manifolds without solving the eigenvalue problem for each spin speed is presented. The performance of the reduced system is analysed in function of the spin speed.

Nomenclature
**C**: damping matrix
**G**: gyroscopic matrix
**M**: mass matrix
**K**: stiffness matrix
**Γ**: nonlinear coefficients matrix
**x**: generalized coordinates
$\alpha$: nonlinearity parameter
$\Omega$: spin speed
$a_{ij}, b_{ij}$: coefficients for the approximation of the manifold
$c_{11}, c_{22}$: elements of the damping matrix
$g_{12}, g_{21}$: elements of the gyroscopic matrix
$m_{11}, m_{22}$: elements of the mass matrix
$k_{11}, k_{22}$: elements of the stiffness matrix
$u$: modal displacement

*v*:         modal speed

## 1   INTRODUCTION

The increasing need of optimized performance of structural systems enhance the importance of the nonlinear effects on their dynamics. To avoid the use of oversized components in obtaining optimal functioning of their systems, the designers and analysts began to extend the linear models to incorporate some kind of elements to reproduce the nonlinear peculiarities.

For linear structures the modal analysis technique is one of the most valuable tools and from its results the response of one structure may be found by solving ordinary differential equations with constant coefficients (Meirovitch, 1996). The key to this technique is to determine the linear transformation that takes the problem from one space where the description of the dynamical behavior is complex (in the sense where couplings between coordinates demand special care to solve the system of differential equations) to another space where the system of differential equations may be readily solved. The modal analysis allowed the development of reduction techniques that are very well developed nowadays. For rotor-bearing systems, perhaps the most simple and known reduction is the pseudo-modal method (Lalanne and Ferraris, 1990). This method of reduction retains the modal base of the system with null speed to describe the motions of the system at several spin speeds, and the reduction is achieved by retaining the low frequencies modes. Another kind of model reduction is the balanced model reduction (Mohiuddin et al, 1998).

The techniques mentioned above are well suited for linear systems. In rotor-bearing system there are many sources of nonlinearities, such as play in bearings, fluid dynamics in journal bearings, contacts between rotor and stator, among others (Yamamoto and Ishida, 2001; Ehrich, 1992; Vance, 1988). These phenomena lead to nonlinear differential equations of motion to express the dynamics of the system. Several methods are available, such as perturbation methods, harmonic balance methods, normal forms and center manifold methods (Nayfeh and Mook, 1979; Nayfeh and Balachandran, 1995; Guckenheimer and Holmes, 1986; Hsu, 1983-a; Hsu, 1983-b; Szemplinska-Stupinicka, 1979-a; Szemplinska-Stupinicka, 1990-b; Yu, 1998; Jézéquel and Lamarque, 1991; Sinou et al., 2003-a; Sinou et al., 2003-b; Sinou et al., 2003-c; Sinou et al., 2004; Huseyin, 2002; Raghothama and Narayanan, 1999 ; Cameron and Griffin, 1989; Nelson and Nataraj, 1989). To try one of the reduction techniques, one approach is to linearize the system near to an equilibrium point. Clearly, this approach is valid when the vibrations are sufficiently small. In addition, it must be noted that the nonlinearities may lead to coordinate couplings that makes the linear modal reduction techniques difficult. Specialized methods for creating reduced models of nonlinear systems are available (Steindl and Troger, 2001), where the reduction is done with the aid of nonlinear Galerkin method and center manifold reduction.

The concept of nonlinear normal modes of vibration presents a great potential in system modeling and reduction techniques. Rosenberg's work (Rosenberg, 1966; Rosenberg, 1962) is the cornerstone of the study of nonlinear normal modes. He defined the nonlinear normal mode for autonomous systems as one synchronous motion with fixed relations between generalized coordinates. In recent years, many studies explored the notion of nonlinear normal modes and nonlinear natural frequencies (Szemplinska-Stupinicka, 1979-a; Szemplinska-Stupinicka, 1990-b; Vakakis et al., 1996).  The invariant manifold approach (Shaw and Pierre, 1991; Shaw and Pierre, 1993) brings the philosophy of the modal analysis to the nonlinear problems. In this approach a nonlinear normal mode is a motion that takes place on an invariant manifold that is tangent to the linear modal subspaces at the point of equilibrium. This definition leads to a nonlinear transformation that relates physical coordinates to nonlinear modal coordinates. The invariant manifold methodology has one aspect that seems to be very promising when searching for reduced models, since one nonlinear normal mode is constructed by projecting the other modes over it by means of a nonlinear relationship. These projections contain the nonlinear effects, and the performance of the reduced model is adapted to weak nonlinearities in virtue of the power series used to obtain the approximation. This approach allowed the incorporation of the nonlinear effects systematically into real world structures (Soares and Mazzilli, 2000).

This paper presents a numerical investigation of a nonlinear rotor-bearing system using the invariant manifold based methodology. As the solution of the eigenproblem of a rotor is dependent on the spin speed and the cost to search for it is very high when the dynamics of the system is to be known for several spin speeds, one strategy is presented to allow the determination of the linear invariant manifolds for several spin speeds based only in a few really calculated ones. By this strategy, a reduced model is constructed and its performance is evaluated and discussed.

## 2 THE NONLINEAR NORMAL MODES

Consider a gyroscopic system with the following equation of motion:

$$\mathbf{M}\ddot{\mathbf{x}} + (\mathbf{C} + \Omega\mathbf{G})\dot{\mathbf{x}} + \mathbf{K}\mathbf{x} + \mathbf{\Gamma}\mathbf{x}^{3*} = \mathbf{0} \tag{1}$$

where $\mathbf{M}$ is the mass matrix (symmetric), $\mathbf{K}$ is the stiffness matrix (symmetric), $\mathbf{C}$ is the damping matrix (symmetric), $\mathbf{G}$ is the gyroscopic matrix (skew symmetric) and $\mathbf{x}$ is the vector of generalized coordinates. The term $\mathbf{\Gamma}\mathbf{x}^{3*}$ represents a stiffness type nonlinearity, observing that 3* means that the power acts over the elements of the vector and $\mathbf{\Gamma}$ is a diagonal matrix of coefficients. The matrices are given by

$$\mathbf{M} = \begin{bmatrix} m_{11} & 0 \\ 0 & m_{22} \end{bmatrix};\ \mathbf{K} = \begin{bmatrix} k_{11} & 0 \\ 0 & k_{22} \end{bmatrix};\ \mathbf{C} = \begin{bmatrix} c_{11} & 0 \\ 0 & c_{22} \end{bmatrix};\ \mathbf{G} = \begin{bmatrix} 0 & -g_{12} \\ g_{21} & 0 \end{bmatrix} \text{ and } \mathbf{\Gamma} = \begin{bmatrix} \alpha & 0 \\ 0 & 0 \end{bmatrix}. \tag{2}$$

Equation (1) may be written in the state space as:

$$\begin{cases} \dot{x}_1 = y_1 \\ \dot{y}_1 = -\dfrac{c_{11}}{m_{11}} y_1 - \dfrac{k_{11}}{m_{11}} x_1 + \Omega \dfrac{g_{12}}{m_{11}} y_2 - \dfrac{\alpha}{m_{11}} x_1^3 \end{cases} \tag{3}$$

$$\begin{cases} \dot{x}_2 = y_2 \\ \dot{y}_2 = -\dfrac{c_{22}}{m_{22}} y_2 - \dfrac{k_{22}}{m_{22}} x_2 - \Omega \dfrac{g_{21}}{m_{22}} y_1 \end{cases} \tag{4}$$

Then, to apply the invariant manifold approach, the following functions can be defined

$$f_1(x_1, x_2, y_1, y_2) = -\dfrac{c_{11}}{m_{11}} y_1 - \dfrac{k_{11}}{m_{11}} x_1 + \Omega \dfrac{g_{12}}{m_{11}} y_2 - \dfrac{\alpha}{m_{11}} x_1^3 \tag{5}$$

$$f_2(x_1, x_2, y_1, y_2) = -\dfrac{c_{22}}{m_{22}} y_2 - \dfrac{k_{22}}{m_{22}} x_2 - \Omega \dfrac{g_{21}}{m_{22}} y_1 \tag{6}$$

The first pair of modal displacement and speed is taken as reference. The second mode is written as function of the first as a fifth order power series expansion:

$$\begin{cases} x_1 = u \\ y_1 = v \end{cases} \quad (7)$$

$$x_2 = X_2(u,v) = a_{10}u + a_{01}v + a_{20}u^2 + a_{11}uv + a_{02}v^2 + a_{30}u^3 + a_{21}u^2v + a_{12}uv^2 + a_{03}v^3 + a_{40}u^4 + \\ + a_{31}u^3v^1 + a_{22}u^2v^2 + a_{13}uv^3 + a_{04}v^4 + a_{50}u^5 + a_{41}u^4v^1 + a_{32}u^3v^2 + a_{23}u^2v^3 + a_{14}uv^4 + a_{05}v^5 \quad (8)$$

$$y_2 = Y_2(u,v) = b_{10}u + b_{01}v + b_{20}u^2 + b_{11}uv + b_{02}v^2 + b_{30}u^3 + b_{21}u^2v + b_{12}uv^2 + b_{03}v^3 + b_{40}u^4 + \\ + b_{31}u^3v^1 + b_{22}u^2v^2 + b_{13}uv^3 + b_{04}v^4 + b_{50}u^5 + b_{41}u^4v^1 + b_{32}u^3v^2 + b_{23}u^2v^3 + b_{14}uv^4 + b_{05}v^5 \quad (9)$$

Equations (8) and (9) are time differentiated to obtain $\dot{X}_2(u,v,\dot{u},\dot{v})$ and $\dot{Y}_2(u,v,\dot{u},\dot{v})$. Then, the time derivatives of $u$ and $v$ are obtained with the aid of equations (3) and (5). Finally, the previous results are inserted in the equations (4), which gives:

$$Y_2(u,v) - \dot{X}_2(u,v, f_1(u, X_2(u,v), v, Y_2(u,v))) = 0 \\ f_2(u, X_2(u,v), v, Y_2(u,v)) - \dot{Y}_2(u,v, f_1(u, X_2(u,v), v, Y_2(u,v))) = 0 \quad (10)$$

Applying the definitions of the functions involved in the equations, one can obtain a set of 40 equations (ignoring the terms of order greater than 5) on the coefficients $a_{ij}$ and $b_{ij}$. These equations may be solved by blocks, that is, for each set of coefficients of a given order there is one set of equations that is function of the previous order coefficients. For the terms of first order, the equations to be solved are represented by the following nonlinear system of equations:

$$\begin{bmatrix} \dfrac{k_{11}}{m_{11}} & 0 & 0 & 1 \\ 0 & -\dfrac{k_{22}}{m_{22}} & \dfrac{k_{11}}{m_{11}} & -\dfrac{c_{22}}{m_{22}} \\ \dfrac{c_{11}}{m_{11}} & -1 & 0 & 1 \\ -\dfrac{k_{22}}{m_{22}} & 0 & \dfrac{c_{11}}{m_{11}} - \dfrac{c_{22}}{m_{22}} & -1 \end{bmatrix} \begin{bmatrix} a_{01} \\ a_{10} \\ b_{01} \\ b_{10} \end{bmatrix} = \begin{bmatrix} \dfrac{\Omega g_{12} a_{01} b_{10}}{m_{11}} \\ \dfrac{\Omega g_{12} b_{01} b_{10}}{m_{11}} \\ \dfrac{\Omega g_{12} a_{01} b_{01}}{m_{11}} \\ \dfrac{\Omega g_{12} b_{01}^2}{m_{11}} \end{bmatrix} \quad (11)$$

The system of equations (11) is nonlinear in the linear coefficients, and its solution is not straightforward. For the underdamped system there are two sets of real roots representing the linear modes of vibration, and the solutions are denoted by $a_{10}^{(1)}$, $a_{01}^{(1)}$, $b_{10}^{(1)}$, $b_{01}^{(1)}$, $a_{10}^{(2)}$, $a_{01}^{(2)}$, $b_{10}^{(2)}$ and $b_{01}^{(2)}$, where the superscript indicates the mode. However, there is an easier way to get the solutions of (10): as the linear coefficients are elements of the eigenvectors of the system expressed in other base than the usual, it is possible to determine them by solving the traditional eigenvalue problem and then perform a normalization in such a way that they represent the modal planar manifolds.

After finding the two sets of coefficients for the linear modes, the terms of higher order may be calculated by solving a linear system of equations. In the problem considered here, all the second and fourth order terms are zero. For the third order terms the linear system consists of an 8×8 coefficient matrix with elements that are functions of the first order terms, the physical constants and the spin speed. Next the fifth order terms are found by solving a 12×12 linear system that is function of the linear and third order terms, the physical constants and the spin speed. The two linear systems can be defined as follow

$$\begin{bmatrix} \theta_3 & 0 & 0 & 0 & \theta_4 & \theta_1 & 0 & 0 \\ 0 & \theta_1 & 0 & 0 & \theta_2 & 0 & 0 & 0 \\ -3 & \theta_5 & 2\theta_1 & 0 & 0 & \theta_2 & 0 & 0 \\ 0 & \theta_3 & 0 & 0 & -3 & \theta_6 & 2\theta_1 & 0 \\ 0 & -2 & 2\theta_5 & 3\theta_1 & 0 & 0 & \theta_2 & 0 \\ 0 & 0 & \theta_3 & 0 & 0 & -2 & \theta_7 & 3\theta_1 \\ 0 & 0 & -1 & 3\theta_5 & 0 & 0 & 0 & \theta_2 \\ 0 & 0 & 0 & \theta_3 & 0 & 0 & -1 & \theta_8 \end{bmatrix} \begin{bmatrix} a_{30} \\ a_{21} \\ a_{12} \\ a_{03} \\ b_{30} \\ b_{21} \\ b_{12} \\ b_{03} \end{bmatrix} = \begin{bmatrix} -\alpha b_{01} \\ -\alpha a_{01} \\ 0 \\ 0 \\ 0 \\ 0 \\ 0 \\ 0 \end{bmatrix} \quad (12)$$

$$\begin{bmatrix} 0 & \theta_1 & 0 & 0 & 0 & 0 & \theta_2 & 0 & 0 & 0 & 0 & 0 \\ \theta_3 & 0 & 0 & 0 & 0 & 0 & \theta_4 & \theta_1 & 0 & 0 & 0 & 0 \\ -5 & \theta_5 & 2\theta_1 & 0 & 0 & 0 & 0 & \theta_2 & 0 & 0 & 0 & 0 \\ 0 & \theta_3 & 0 & 0 & 0 & 0 & -5 & \theta_6 & 2\theta_1 & 0 & 0 & 0 \\ 0 & -4 & 2\theta_5 & 3\theta_1 & 0 & 0 & 0 & 0 & \theta_2 & 0 & 0 & 0 \\ 0 & 0 & \theta_3 & 0 & 0 & 0 & 0 & -4 & \theta_7 & 3\theta_1 & 0 & 0 \\ 0 & 0 & -3 & 3\theta_5 & 4\theta_1 & 0 & 0 & 0 & 0 & \theta_2 & 0 & 0 \\ 0 & 0 & 0 & \theta_3 & 0 & 0 & 0 & 0 & -3 & \theta_8 & 4\theta_1 & 0 \\ 0 & 0 & 0 & -2 & 4\theta_5 & 5\theta_1 & 0 & 0 & 0 & 0 & \theta_2 & 0 \\ 0 & 0 & 0 & 0 & \theta_3 & 0 & 0 & 0 & 0 & -2 & \theta_9 & 5\theta_1 \\ 0 & 0 & 0 & 0 & -1 & 5\theta_5 & 0 & 0 & 0 & 0 & 0 & \theta_2 \\ 0 & 0 & 0 & 0 & 0 & \theta_3 & 0 & 0 & 0 & 0 & -1 & \theta_{10} \end{bmatrix} \begin{bmatrix} a_{50} \\ a_{41} \\ a_{32} \\ a_{23} \\ a_{14} \\ a_{05} \\ b_{50} \\ b_{41} \\ b_{32} \\ b_{23} \\ b_{14} \\ b_{05} \end{bmatrix} = \begin{bmatrix} -a_{21}\alpha + \dfrac{\Omega g_{12} a_{21} b_{30}}{m_{11}} \\ -b_{21}\alpha + \dfrac{\Omega g_{12} a_{21} b_{30}}{m_{11}} \\ -2a_{12}\alpha + \dfrac{\Omega g_{12} a_{21} b_{21} + 2\Omega g_{12} a_{21} b_{21}}{m_{11}} \\ -2b_{12}\alpha + \dfrac{2\Omega g_{12} b_{12} b_{30} + \Omega g_{12} b_{21}^2}{m_{11}} \\ -3a_{03}\alpha + \dfrac{\Omega g_{12} a_{21} b_{12} + 3\Omega g_{12} a_{03} b_{30} + 2\Omega g_{12} a_{12} b_{21}}{m_{11}} \\ -3b_{03}\alpha + \dfrac{3\Omega g_{12} b_{21} b_{12} + 3\Omega g_{12} b_{03} b_{30}}{m_{11}} \\ \dfrac{\Omega g_{12} a_{21} b_{03} + 3\Omega g_{12} a_{03} b_{21} + 2\Omega g_{12} a_{12} b_{12}}{m_{11}} \\ \dfrac{4\Omega g_{12} b_{03} b_{21} + 2\Omega g_{12} b_{12}^2}{m_{11}} \\ \dfrac{2\Omega g_{12} a_{12} b_{03} + 3\Omega g_{12} a_{03} b_{12} + 3\Omega g_{12} a_{13} b_{02}}{m_{11}} \\ \dfrac{5\Omega g_{12} b_{03} b_{12}}{m_{11}} \\ \dfrac{3\Omega g_{12} a_{03} b_{03}}{m_{11}} \\ \dfrac{3\Omega g_{12} b_{03}^2}{m_{11}} \end{bmatrix} \quad (13)$$

where :

$$\begin{aligned}
&\theta_1 = \frac{k_{11}}{m_{11}} - \frac{\Omega g_{12} b_{10}}{m_{11}}; & &\theta_2 = -\frac{\Omega g_{12} a_{01}}{m_{11}} + 1; \\
&\theta_3 = -\frac{k_{22}}{m_{22}}; & &\theta_4 = -\frac{\Omega g_{12} b_{01}}{m_{11}} - \frac{c_{22}}{m_{22}}; \\
&\theta_5 = \frac{c_{11}}{m_{11}} - \frac{\Omega g_{12} b_{01}}{m_{11}}; & &\theta_6 = -\frac{2\Omega g_{12} b_{01}}{m_{11}} - \frac{c_{22}}{m_{22}} + \frac{c_{11}}{m_{11}}; \\
&\theta_7 = -\frac{3\Omega g_{12} b_{01}}{m_{11}} - \frac{c_{22}}{m_{22}} + \frac{2c_{11}}{m_{11}}; & &\theta_8 = -\frac{4\Omega g_{12} b_{01}}{m_{11}} - \frac{c_{22}}{m_{22}} + \frac{3c_{11}}{m_{11}}; \\
&\theta_9 = -\frac{5\Omega g_{12} b_{01}}{m_{11}} - \frac{c_{22}}{m_{22}} + \frac{4c_{11}}{m_{11}}; & &\theta_{10} = -\frac{6\Omega g_{12} b_{01}}{m_{11}} - \frac{c_{22}}{m_{22}} + \frac{5c_{11}}{m_{11}}.
\end{aligned} \qquad (14)$$

The linear systems are solved for the two sets of real solutions of (11), giving the nonlinear coefficients for each mode. It is interesting to note that when the spin speed is zero all the coefficients of equation (8) and equation (9) are zero, which is expected since the system is coupled by the gyroscopic matrix. After finding all the constants of the equations (6), they are inserted in the equations (3) to obtain the projection of the dynamics of each mode over $u$ and $v$, described by the following oscillator:

$$\begin{cases} \dot{u}_1 = v_1 \\ \dot{v}_1 = -\frac{c_{11}}{m_{11}} v_1 - \frac{k_{11}}{m_{11}} u_1 + \Omega \frac{g_{12}}{m_{11}} Y_2(u_1, v_1) - \alpha u_1^3 \end{cases} \qquad (15)$$

## 3 STRATEGY FOR GYROSCOPIC SYSTEMS

Generally speaking, the methodology presented starts by finding the planar invariant manifolds representing the linear normal modes of the linearized system, and then looking for the nonlinear invariant manifolds tangent to the linear ones that accounts for the nonlinearities.

For a gyroscopic system it may be a problem since the planar invariant manifolds are function of the spin speed. It means that one must perform at least a modal analysis for each value of the spin speed and then solve the linear systems for the higher-order terms, two systems for each order. Things get worse when the number of degrees of freedom of the system grows, because the size of the matrices involved in the calculations grows dramatically. One simple strategy to avoid such an enormous numerical effort is to find the first order coefficients for relatively few values of the spin speed and approximate the relationship between the coefficients and the spin speed by least squares polynomials. This strategy will be illustrated by a numerical example in the next section.

### 3.1 EXAMPLE

This example consists of a rotor-bearing system illustrated in Figure 1. The shaft is 0.4m long and its diameter is 0.02m. The disk has a diameter of 0.55 m and thickness of 0.05m. They are both made of steel with $E=2\times10^{11}$ N/m$^2$ and $\rho$=7800 kg/m$^3$. The bearings are asymmetric with $k_{xx}$=5×10$^5$ N/m and $k_{zz}$=2.1 N/m. The stiffness in the $X$ direction has a cubic nonlinearity associated. The position of the disk and the bearing is $l_D$=0.13m and $l_P$=0.27m, respectively. The physical characteristics of the system lead to the following numerical values for the matrices of the equation of motion, obtained by a Rayleigh-Ritz procedure (Lalanne and Ferraris, 1990): $m_{11}=m_{22}$=97.2 kg, $k_{11}$=1.6×10$^6$ N/m, $k_{22}$=2.8×10$^6$ N/m, $c_{11}=c_{22}$=75 Ns/m, $g_{12}=g_{21}$=54.3 Ns/m and $\alpha$=1×10$^{10}$ N/m$^3$. The Campbell's diagram is constructed by solving the eigenvalue problem of the corresponding linear system and is shown in Figure 2. From the diagram, the first mode is at 18.06 Hz and the second mode is at 36.20 Hz. If the inverse whirl is multiplied by three, it will cross the direct whirl curve at approximately 47 Hz, as illustrated in Figure 2. This frequency corresponds to an internal resonance.

In this study the spin speed of the system is allowed to assume any positive value up to 50 Hz. To apply the strategy presented in the previous section, the eigenvalue problem is solved for twenty equally spaced values of the spin speed to find the first order coefficients, as illustrated in Figure 3. The function chosen to interpolate the data presented in the Figure 3 is a 5$^{th}$ degree polynomials. With the interpolated values, the third order coefficients are calculated and shown in Figures 4 and 5. From these Figures it can be seen that there is a singularity at approximately 47 Hz, the frequency of the internal resonance. A result like this is expected since the methodology employed here is not able to deal with systems with internal resonance.

For a spin speed equal to $\Omega = 8$ Hz, the constants of equation (7) are calculated by using the previously presented strategy. For an initial condition completely in the first mode, the dynamics of the system can be described only by the oscillator corresponding to the first mode, which is represented by:

$$\begin{cases} \dot{u}_1 = v_1 \\ \dot{v}_1 = -7.7157 \times 10^{-1} v_1 - 1.6156 \times 10^4 u_1 + 2.7793 \times 10^1 Y_2(u_1, v_1) - 1 \times 10^{10} u_1^3 \end{cases} \quad (15)$$

where the function $Y_2(u_1, v_1)$ is expressed by the equation (9), with the constant defined previously. After integration, equation (8) gives the speed and displacement of the second mode.

The initial conditions for equation (11) are chosen to be: $\begin{bmatrix} u_1 & v_1 \end{bmatrix}_{ini}^T = \begin{bmatrix} 9 \times 10^{-4} & 0 \end{bmatrix}^T$. These initial conditions are sufficient to show the nonlinear effects. The final time for the time integration is chosen to be 0.5s. The results of the simulations for the exact, reduced and linear systems are shown in Figures 6 and 7; the time integration of equation (15) is named as "reduced system" and is compared with the direct integration of the full model (defined by the equations (3) and (4) and named as "exact system"), and with the linear system given by the equations (2) and (3) without the nonlinear term and named as "linear system". The initial conditions for the exact model are obtained from the reduced system and equations (7) and (8) at time $t$=0 s. The comparison of the reduced system with the direct integration shows a very good agreement. It shows that nonlinear effects are present and are well captured by the reduced model. Moreover, Figures 8 and 9 show the modal displacement and speed manifold for the first mode and the results of the associated simulations, respectively.

Then, the same type of simulation was made for $\Omega = 24$ Hz as illustrated in Figures 10-13. In this case, the results given by the reduced model are quite good and the effects of the nonlinear terms appear to be significant. Finally, simulations are made for $\Omega = 44$ Hz, as illustrated in Figure 14-17. In this last case, the nonlinear terms are very important. The modal displacement and speed manifold for the first mode is very complex as illustrated in Figures 16 and 17. Although the comparison between the reduced and exact systems is not perfect, it may be observed that the reduced system is a first approximation of the exact system. Moreover, it shows that the performance of the reduced model near the internal resonance is penalized, as the interaction between the modes cannot be captured.

To investigate the performance of the reduced model over the spin speed range, the correlation between the time response of the reduced model and the time response of the full model is calculated and shown in the Figure 18. It can be seen that the reduced model has a very good performance up to 42 Hz, where the correlation indicates that the coupling between the first and second modes starts to get significant. Moreover, the simulation was carried out for a reduced model constructed with third order expansions for the manifolds and the correlation is plotted in Figure 18. It is shown that the third order expansion gives almost the same results as that of the fifth order expansion up to 32 Hz and then it drops more dramatically than does the fifth order model. The disturbance caused by the internal resonance is more remarkable for the third order reduced model.

## 4 CONCLUSION

The methodology of invariant manifolds for the construction of nonlinear normal modes was employed in a rotor system with a nonlinear cubic stiffness in one plane. It allows the uncoupling of nonlinear problem as in traditional modal analysis. Unfortunately the manifolds need to be calculated for each value of the spin speed, which can be a costly task. To overcome this problem a simple strategy is presented. Firstly, the linear modal manifolds for a relatively few spin speeds must be found. Secondly, a least squares polynomial which describes the behavior of the linear modal manifolds over the spin speed range is adjusted to the points obtained. This description is much more simple to calculate than the original one. It was found that the system under study had an internal resonance. Singularities were observed in the calculation of higher order terms of the manifolds since the invariant manifold methodology employed can not treat this phenomenon.

The order of the power series expansion for the manifolds is a delicate question. In this study, we showed a comparison of the performance of a third order manifold with a fifth order manifold for a given initial condition. However, the order of the series expansion has to be increased if the nonlinear effects are more pronounced of if the system is vibrating near of a internal resonance. This last working condition was explored in the numerical example, where it was shown that the effects of the interaction between the first and the second modes are more pronounced in the case of a third order power series expansion for the manifolds.

For systems with a greater number of degrees of freedom, the interpolation strategy used for the first order coefficients combined with a methodology which takes into account internal resonances can be very interesting.

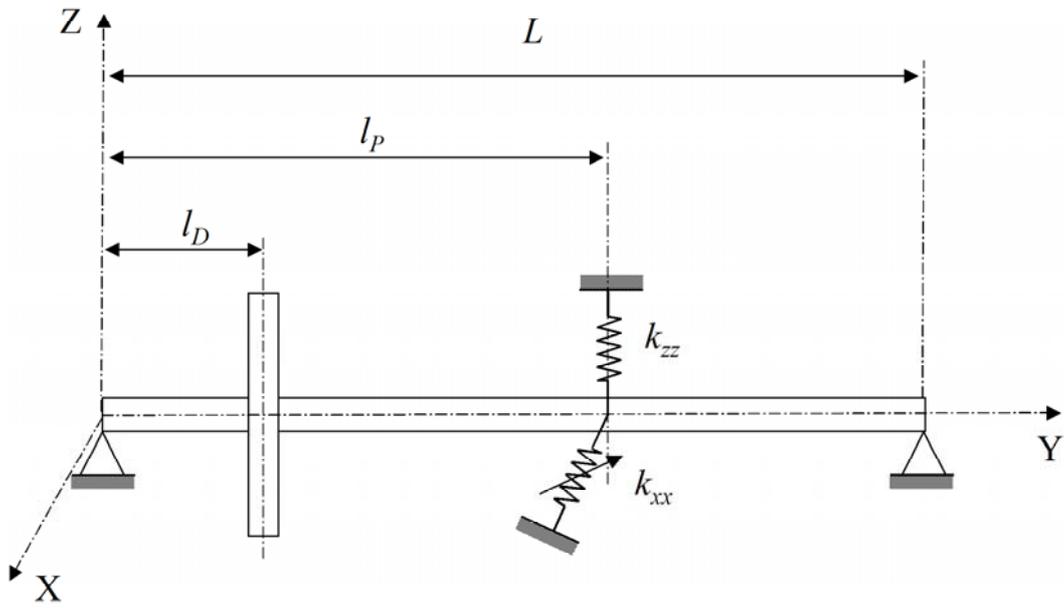

Figure 1 - Rotor with a non-linear bearing

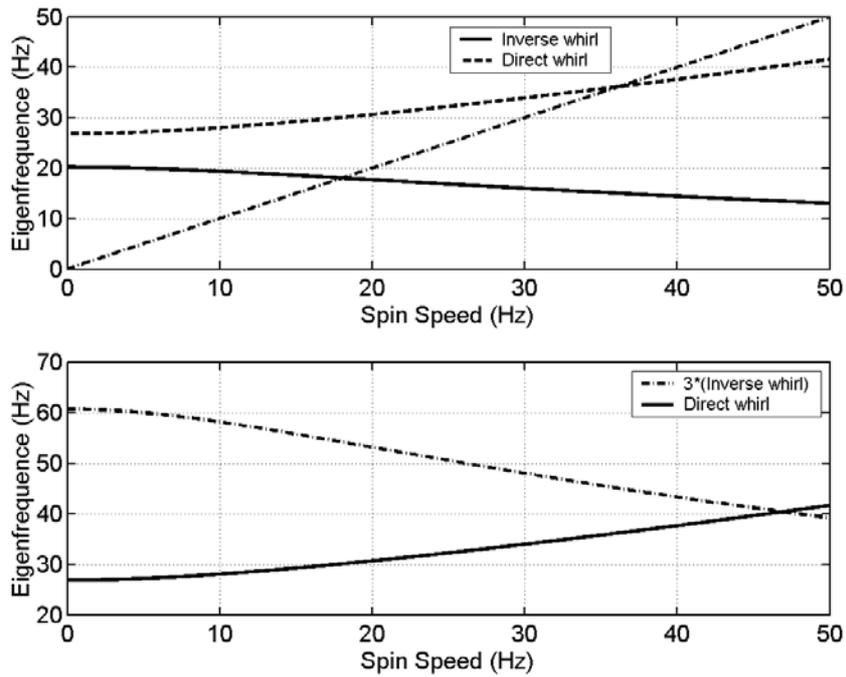

Figure 2 - Campbell's Diagram and a internal resonance frequency

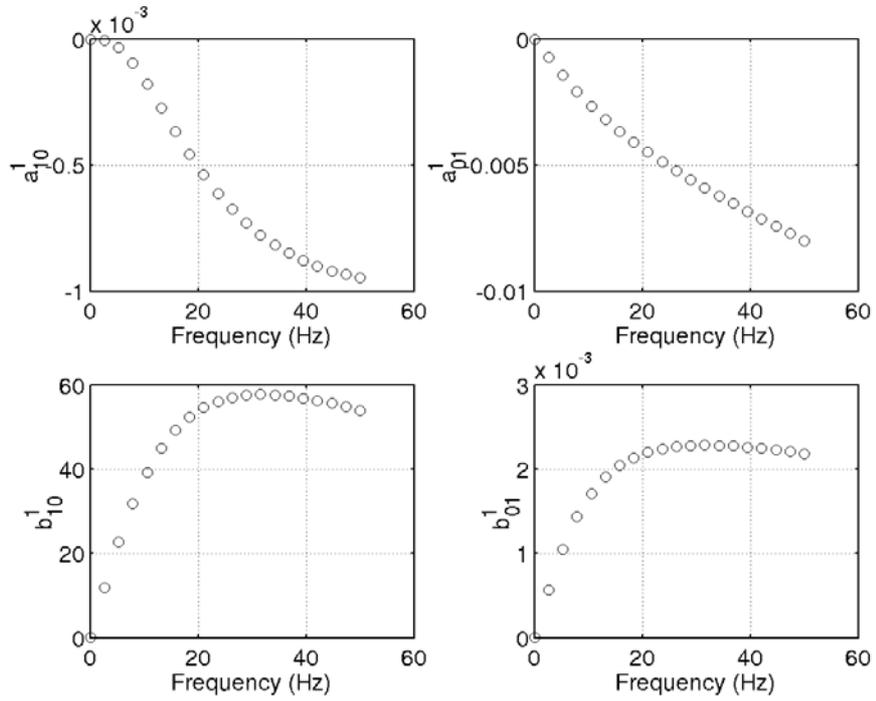

Figure 3- First order coefficients calculated at 20 frequencies

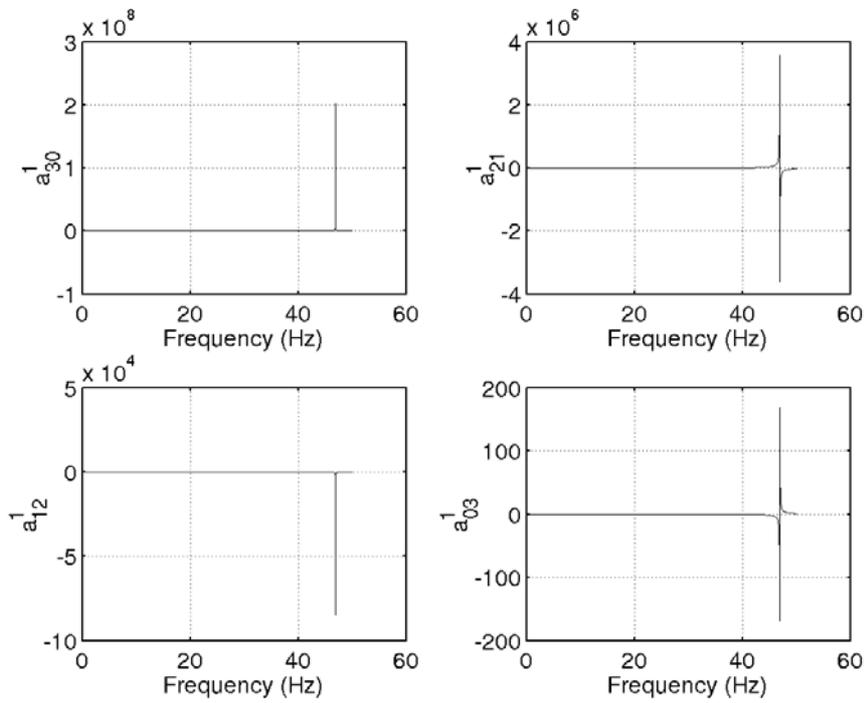

Figure 4 - The third order coefficients ($a_{ij}$) in function of the spin speed, for the first mode

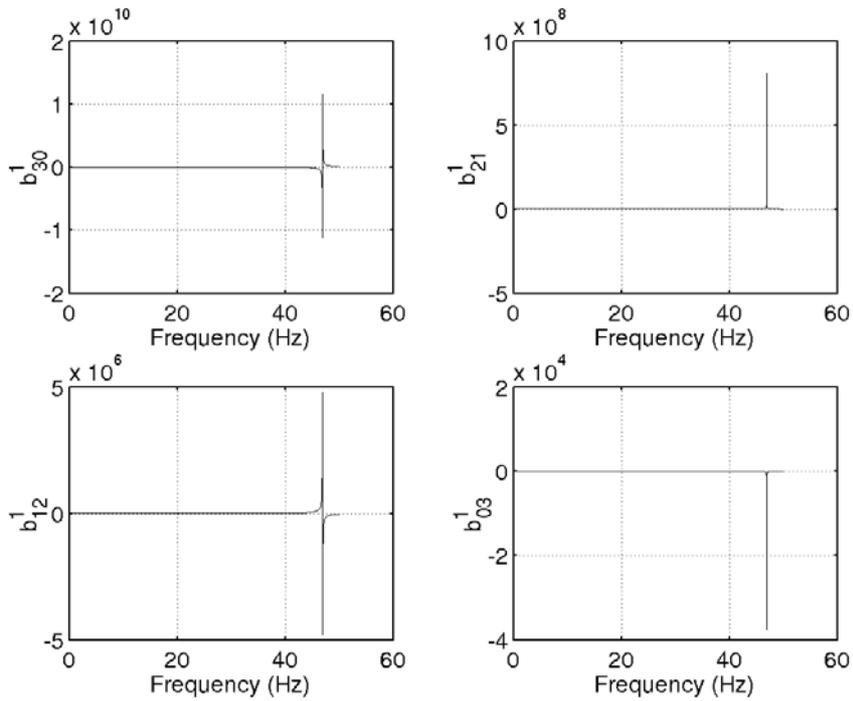

Figure 5 - The third order coefficients ($b_{ij}$) in function of the spin speed, for the first mode

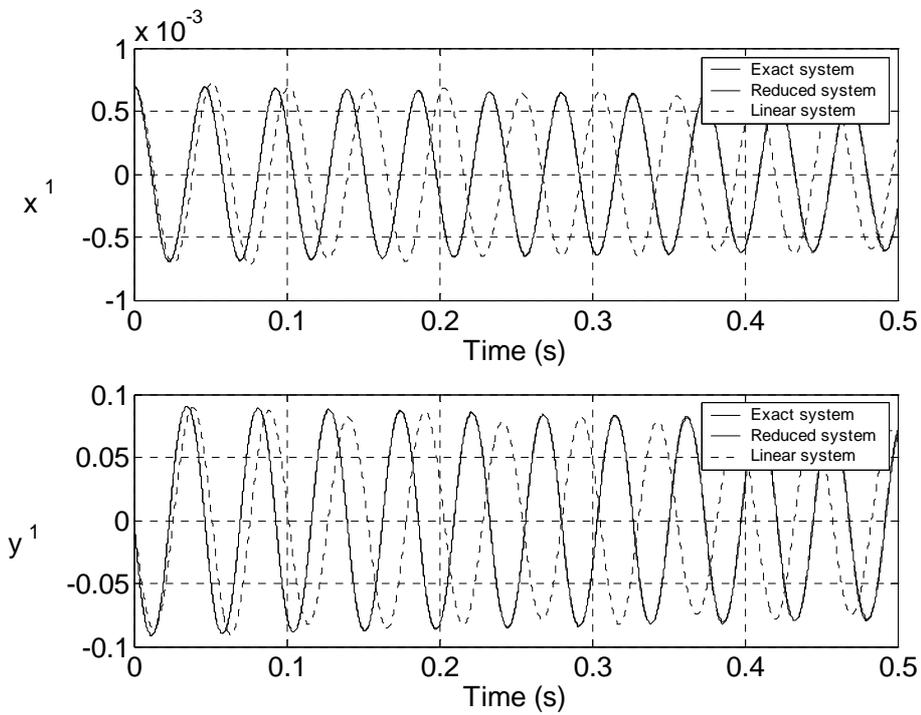

Figure 6 - Comparison between the exact, reduced and linear systems at $\Omega=8$ Hz

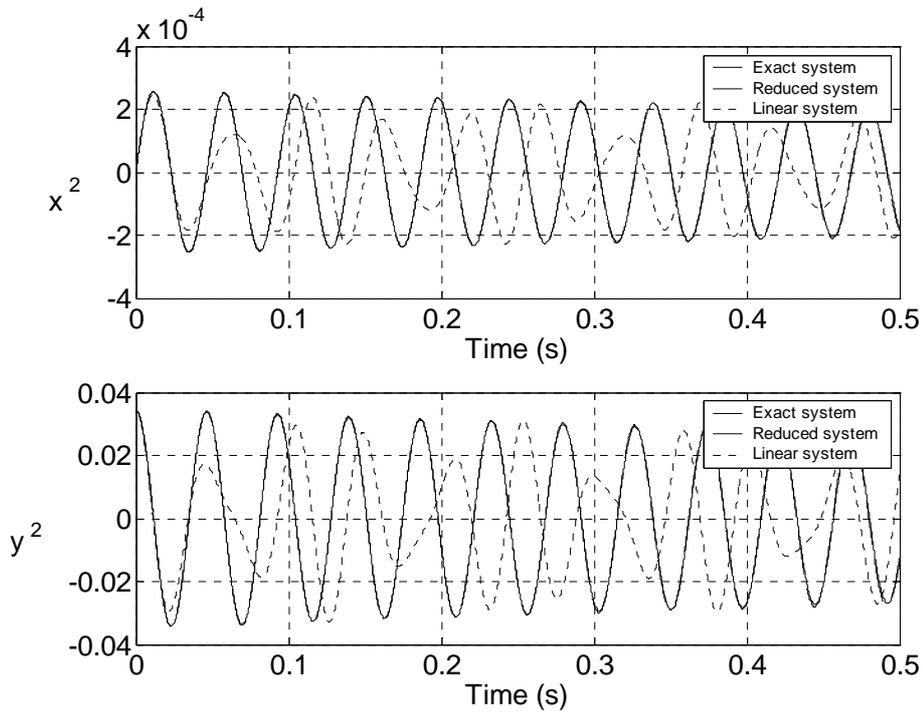

Figure 7 - Comparison between the exact, reduced and linear systems at $\Omega=8$ Hz

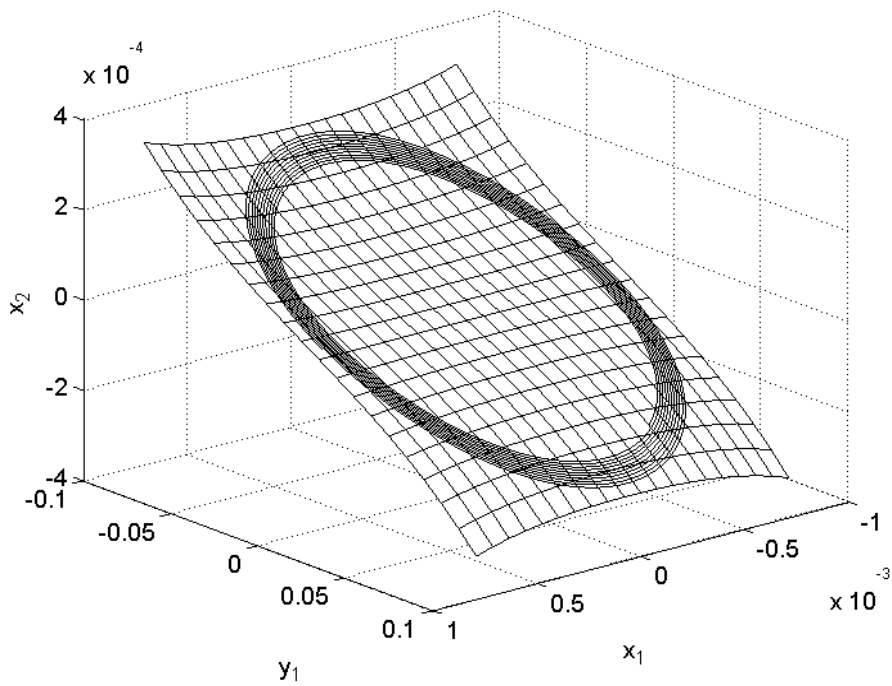

Figure 8 - Modal speed manifold for the first mode at $\Omega=8$ Hz

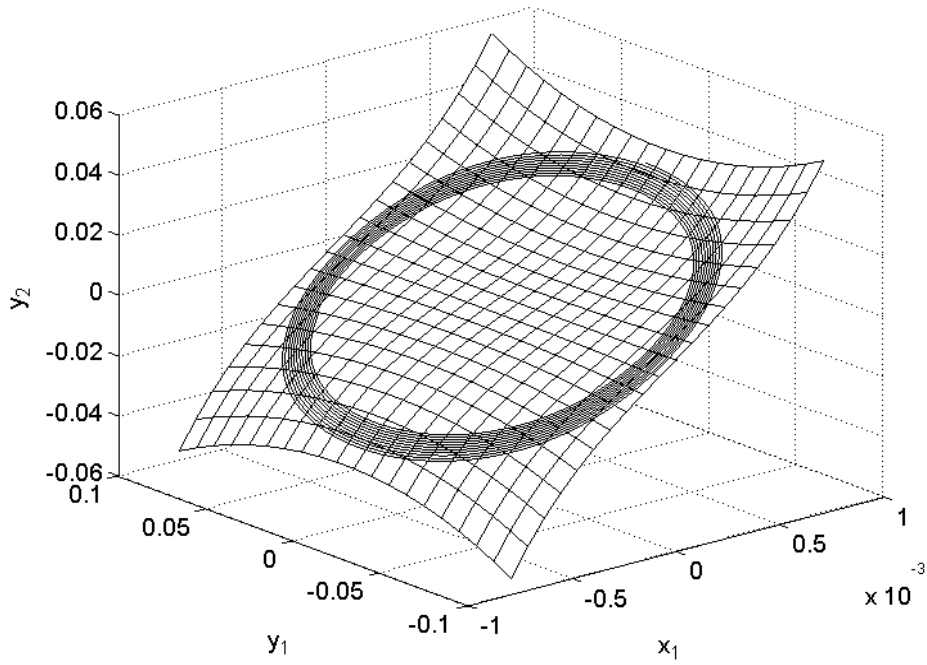

Figure 9 - Modal speed manifold for the first mode at Ω=8 Hz

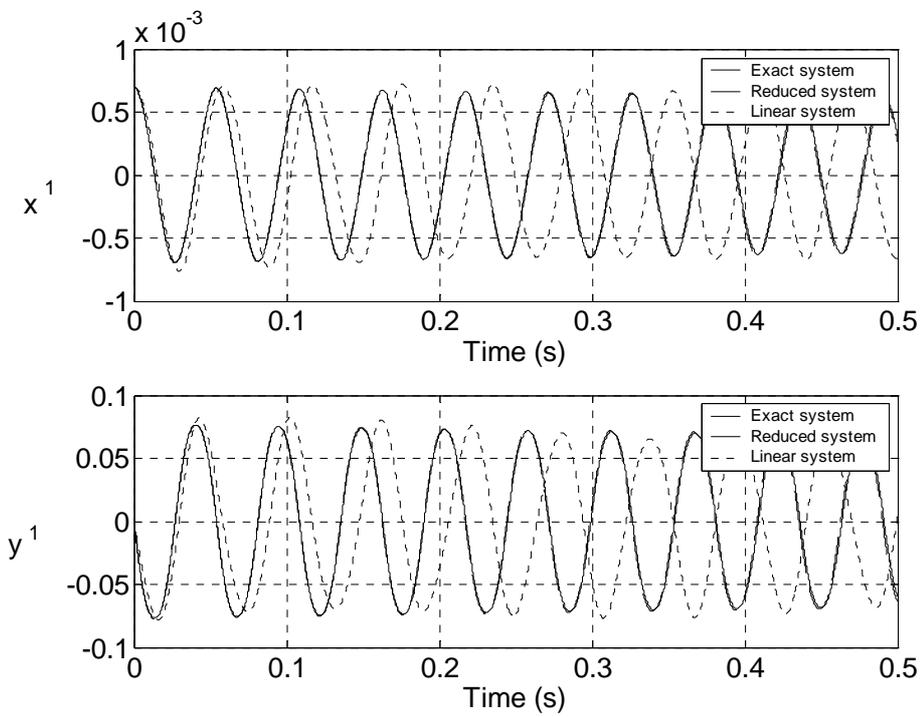

Figure 10 - Comparison between the exact, reduced and linear systems at Ω=24 Hz

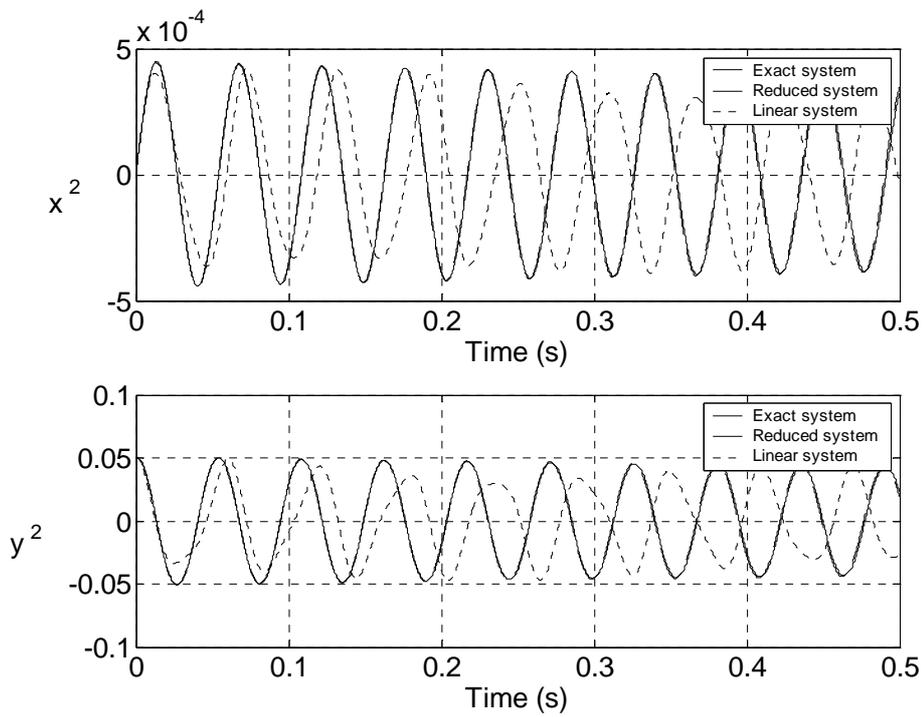

Figure 11 - Comparison between the exact, reduced and linear systems $\Omega=24$ Hz

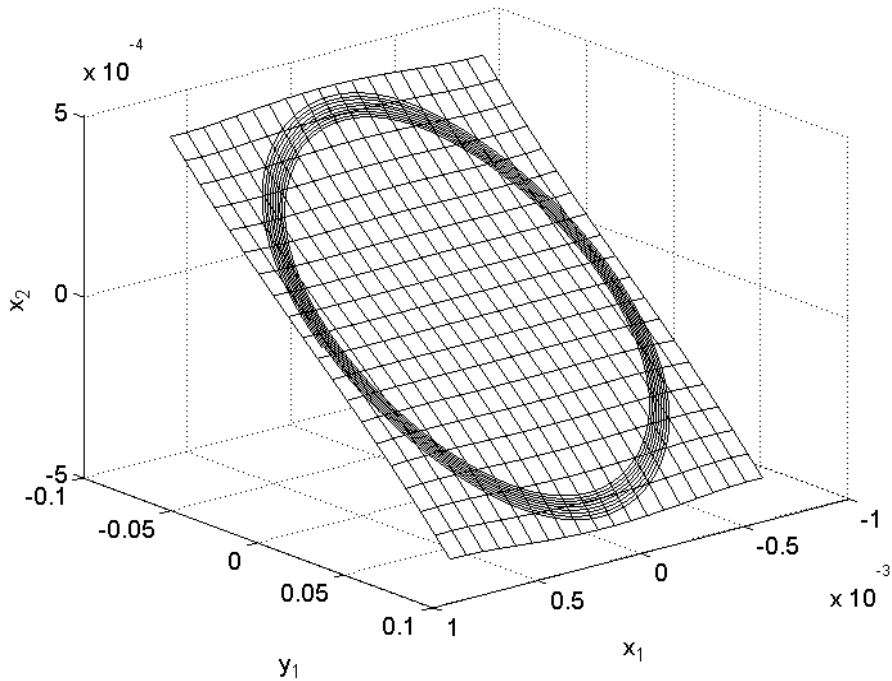

Figure 12 - Modal speed manifold for the first mode for $\Omega=24$ Hz

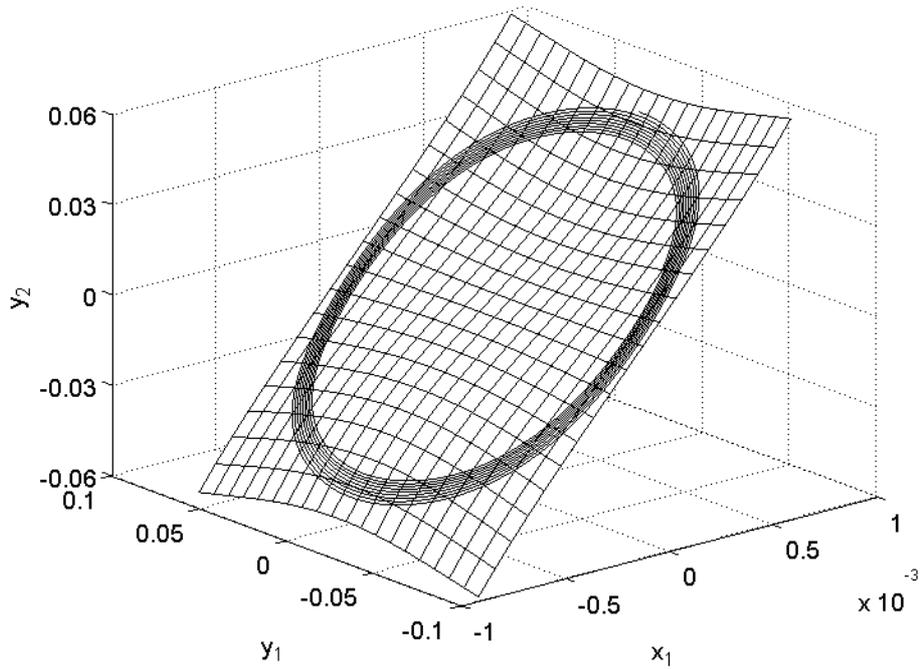

Figure 13 - Modal speed manifold for the first mode for $\Omega=24$ Hz

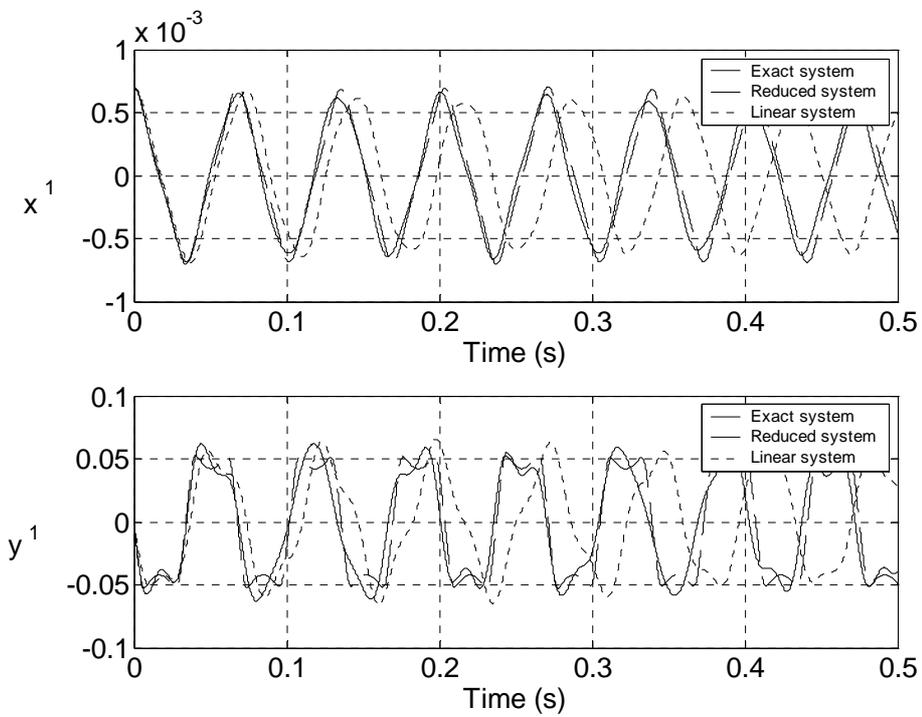

Figure 14 - Comparison between the exact, reduced and linear systems $\Omega=44$ Hz

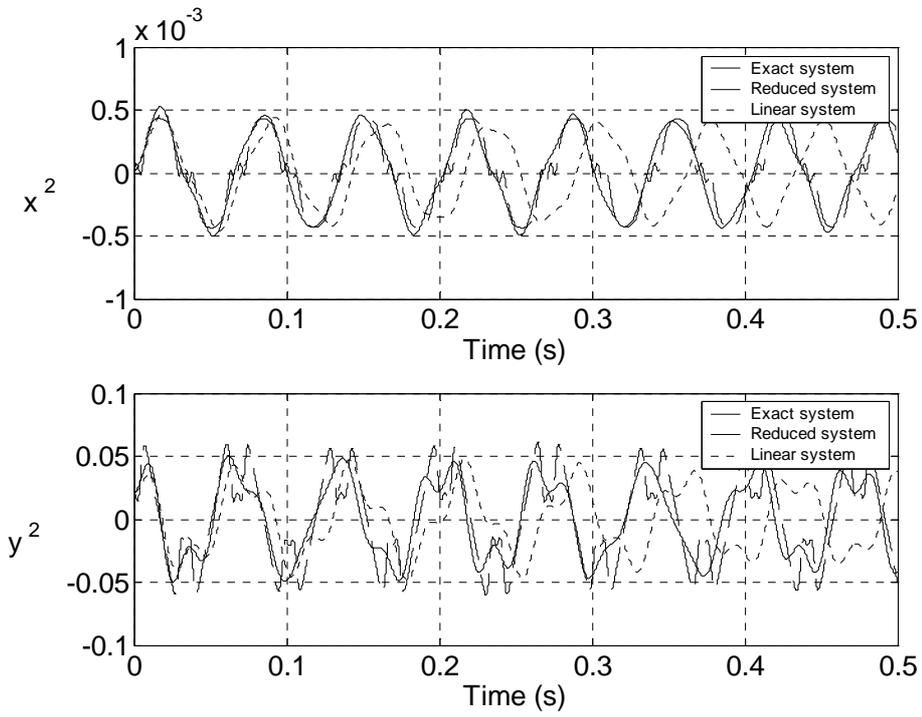

Figure 15 - Comparison between the exact, reduced and linear systems $\Omega=24$ Hz

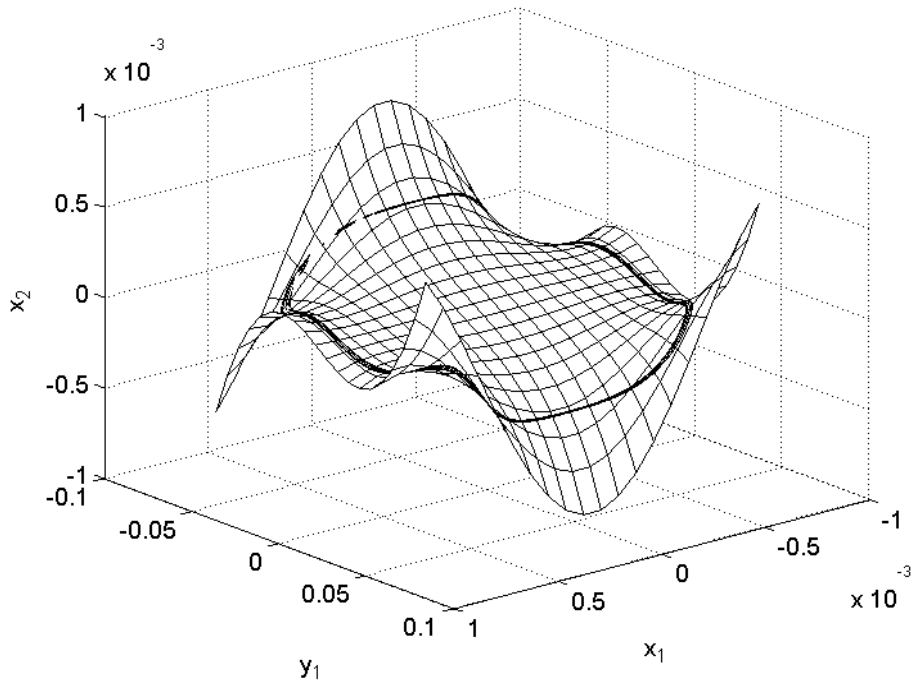

Figure 16 - Modal speed manifold for the first mode for $\Omega=44$ Hz

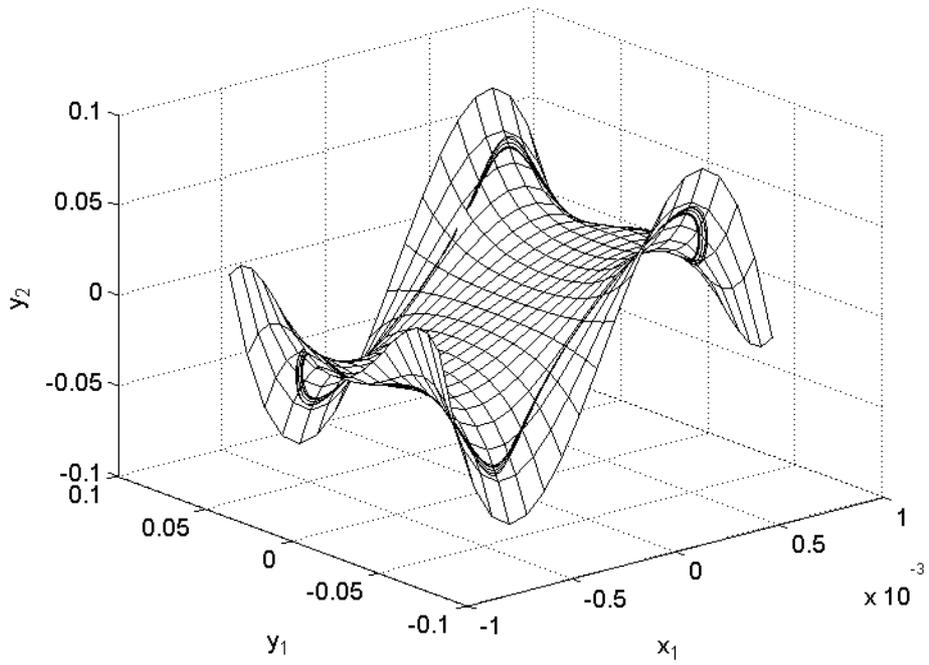

Figure 17 - Modal speed manifold for the first mode for Ω=44 Hz

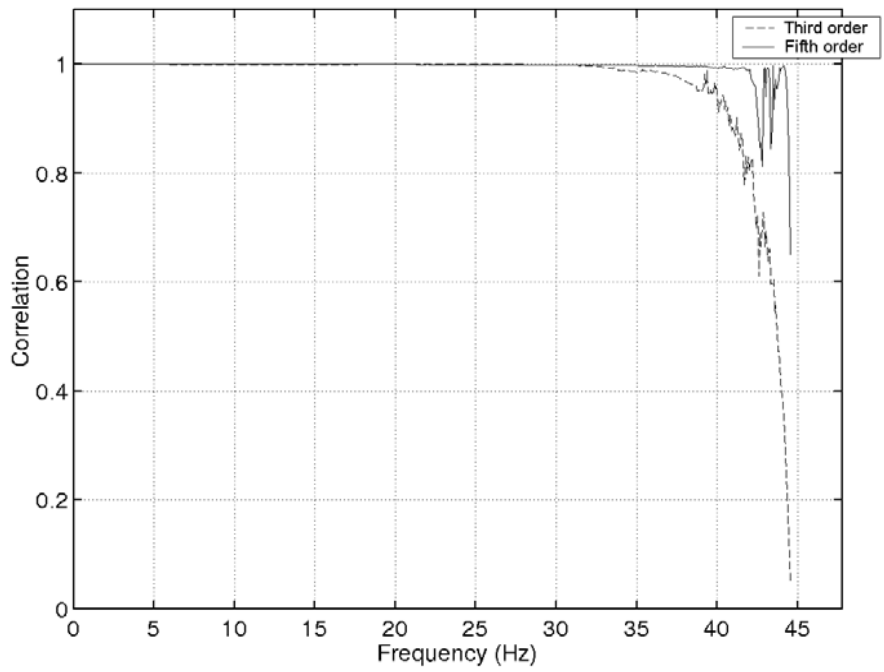

Figure 18 - Correlation between the exact and reduced models in function of the spin speed